\documentstyle[11pt,newpasp,twoside,epsf]{article}
\begin{document}
\title{ Lithium abundances in main-sequence F stars and sub-giants }
\author{Jose Dias do Nascimento Jr, Sylvie Th\'eado and Sylvie Vauclair}
\affil{Laboratoire d'Astrophysique, 14 av. Ed. Belin, 31400 Toulouse,
  France}

\begin{abstract}
The application to main-sequence stars of the rotation-in-duced mixing
theory in the presence of $\displaystyle \mu $-gradients 
leads to partial mixing in the
lithium destruction region, not visible in the atmosphere. The induced
lithium depletion becomes visible in the sub-giant phase as soon as the
convective zone deepens enough. This may explain why the observed ``
lithium dilution " is smoother and the final dilution factor larger than
obtained in standard models, while the lithium abundance variations are very small on the
main sequence.
\end{abstract}

The observations of lithium in main-sequence stars on the hot side of the 
``Boesgaard dip" 
show a very small dispersion for normal stars while a light depletion 
(by a factor 3) is observed in Am stars (Burckhart and
Coupry 2000).
On the other hand, on the sub-giant branch, these stars present a lithium depletion
larger than that predicted by the standard model (do Nascimento et al. 1999). 
These observations 
suggest that, while on the main sequence, the stars suffer in their
internal layers a lithium destruction larger
than the standard one : this extra-destruction, which must not appear at the surface in the
main-sequence phase, is then dredged up during the subsequent evolution on the sub-giant
branch (Vauclair 1991)

It has been suggested several times 
that the process responsible for this extra-depletion could be the result of
rotation-induced mixing.
Computations including such macroscopic motions as described by Zahn 1992 and
Maeder \& Zahn 1998 have recently been performed by Charbonnel and Talon 1999 and 2000.
They show that the observations on the sub-giant branch can nicely be reproduced by such
rotation-induced mixing.
In their computationsi however, the effect of the microscopic diffusion of lithium was not
introduced on the main-sequence, for the reason that in these stars the radiative
acceleration may balance the lithium gravitational settling. 
For helium, on the contrary, the radiative
acceleration is negligible : helium settling was then introduced but not taken into account
while computing the meridional circulation velocity.

As shown by Mestel 1953, Maeder and Zahn 1998, Vauclair 1999, (see also Vauclair 2000 and
Th\'eado and Vauclair 2000),
in the presence of vertical $\displaystyle \mu $-gradients,
the circulation velocity is the sum of two terms which leed to motions in the opposite
direction, one which does not
depend on $\displaystyle \mu $ (the so-called ``$\displaystyle \Omega
$ currents'') and one which gathers the $\displaystyle \mu $
dependent terms
(the ``$\displaystyle \mu $ currents'').
In case of helium gravitational settling, a ``$\displaystyle \mu $ gradient'' builts up
which soon counteracts the standard meridional circulation and an equilibrium situation 
may be reached, which could account for the fact that lithium is preserved on the main
sequence, while extra-mixing occurs below the ``frozen layer".

\begin{figure}
\plotfiddle{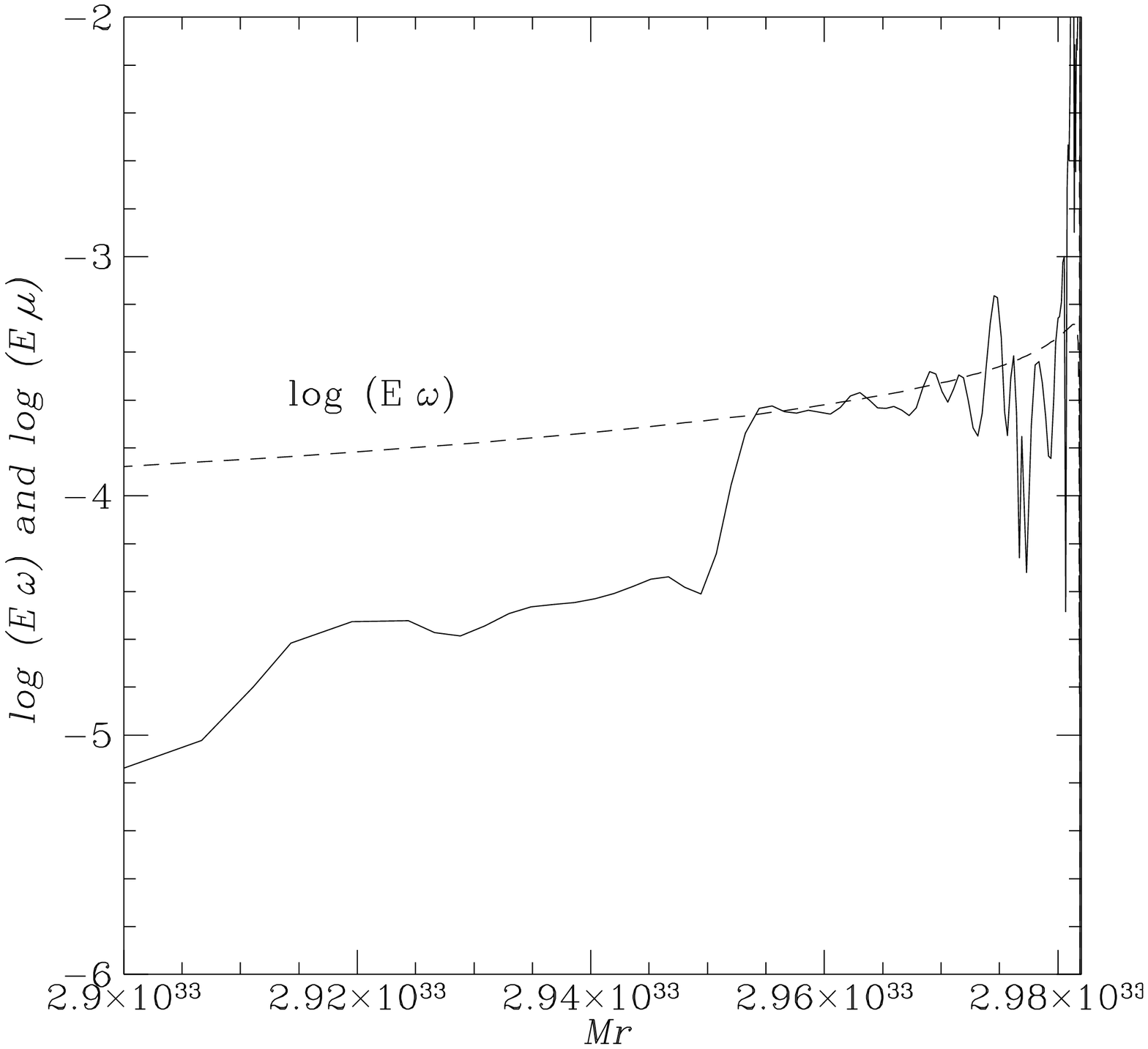}{5.0cm}{0}{30}{30}{-200}{-60}
\plotfiddle{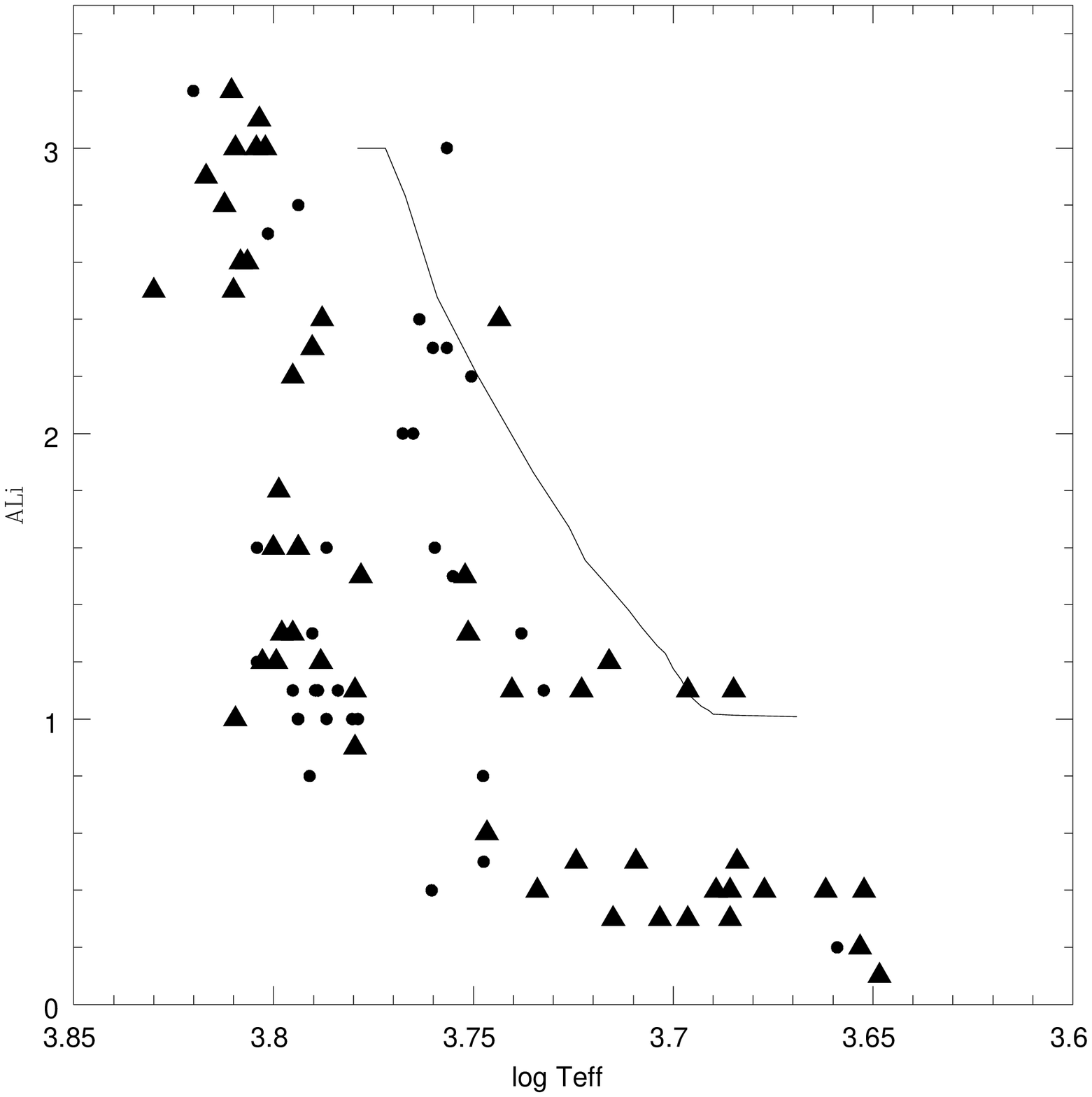}{0cm}{0}{30}{30}{10}{-35}
\caption{\small {computations of the $\Omega$-currents and $\mu $-currents in a $1.5
M_{\odot}$ star
 with a rotation velocity of 40 km.s-1
and lithium evolution on the sub-giant branch 
obtained in this case, compared to the observations}}
\end{figure}

In the present paper, we have computed the 
evolution of a  $1.5 M_{\odot}$ star taking into account the same effects as discussed in
Th\'eado and Vauclair 2000. 
We show that, when the opposite currents are taken into account, the layer just below the
convection zone freezes out while mixing proceeds below. While evolving out of the
main-sequence,  dilution induced by the deepening of the convective zone leeds to a larger
depletion than predicted by the standard model, reproducing the upper envelope of the
observations. More computations are underway to extend these results to other masses and
rotation parameters.

\end{document}